# An Improved Model for Diabetic Retinopathy Detection by using Transfer Learning and Ensemble Learning

[1] Md. Simul Hasan Talukder; Electrical and Electronic Engineering; Rajshahi University of Engineering and Technology Rajshahi, Bangladesh; Email: simulhasantalukder@gmail.com

[2]Ajay Kirshno Sarkar; Electrical and Electronic Engineering; Rajshahi University of Engineering and Technology Rajshahi, Bangladesh; Email: aksarkar@eee.ruet.ac.bd

[3] Sharmin Akter; Biomedical Engineering; Jashore University of Science and Technology, Jashore, Bangladesh; Email: sharmintalukder120@gmail.com

[4] Md. Nuhi-Alamin; Electrical and Electronic Engineering;Rajshahi University of Engineering and TechnologyRajshahi, Bangladesh
Email: nuhialamin@eee.ruet.ac.bd

**Correponding Author:** Md. Simul Hasan Talukder; Electrical and Electronic Engineering; Rajshahi University of Engineering and Technology Rajshahi, Bangladesh; Email: simulhasantalukder@gmail.com

***Abstract***— Diabetic Retinopathy (DR) is an ocular condition caused by a sustained high level of sugar in the blood, which causes the retinal capillaries to block and bleed, causing retinal tissue damage. It usually results in blindness. Early detection can help in lowering the risk of DR and its severity. The robust and accurate prediction and detection of diabetic retinopathy is a challenging task. This paper develops a machine learning model for detecting Diabetic Retinopathy that is entirely accurate. Pre-trained models such as ResNet50, InceptionV3, Xception, DenseNet121, VGG19, NASNetMobile, MobileNetV2, DensNet169, and DenseNet201 with pooling layer, dense layer, and appropriate dropout layer at the bottom of them were carried out in transfer learning (TL) approach. Data augmentation and regularization was performed to reduce overfitting. Transfer Learning model of DenseNet121, Average and weighted ensemble of DenseNet169 and DenseNet201 TL architectures contribute individually the highest accuracy of 100%, the highest precision, recall, F-1 score of 100%, 100%, and 100%, respectively.

## 1. Introduction

Diabetes, commonly known as diabetes mellitus, is a condition in which the human body produces excess blood glucose [1]. It is a universal chronic disease that has been identified as the fourth leading cause of death. [2]. Diabetes has been related to a number of illnesses, including nerve damage, heart disease, stroke, foot difficulties, gum disease, and more [1]. Diabetes is anticipated to affect 336 million people globally, according to the International Diabetes Federation (IDF), with a 7.7% increase expected by 2030 [3, 4]. Diabetic Retinopathy (DR) is a diabetic condition in which the retinal blood vessels enlarge and spill fluid and blood [5]. According to the Mayo Clinic [6], frequent symptoms of DR include visual spots, color impairment, blurred or fluctuating vision, and, in severe cases, complete vision loss in one or both eyes. Long-term high blood sugar levels cause blockage in the retina's micro-vessels, which are critical for nourishing the retina tissues. As a result, the eye strives to create new arteries to provide the retina with the nutrition and oxygen it requires; however, these newly formed vessels are weak and prone to blood loss, resulting in a retinal hemorrhage [7]. In both type 1 and type 2 diabetics, it is a significant cause of blindness [8]. Type-2 diabetes accounts for the majority of diabetes cases [9]. Figure 1 depicts the normal and DR-affected retinas, respectively.

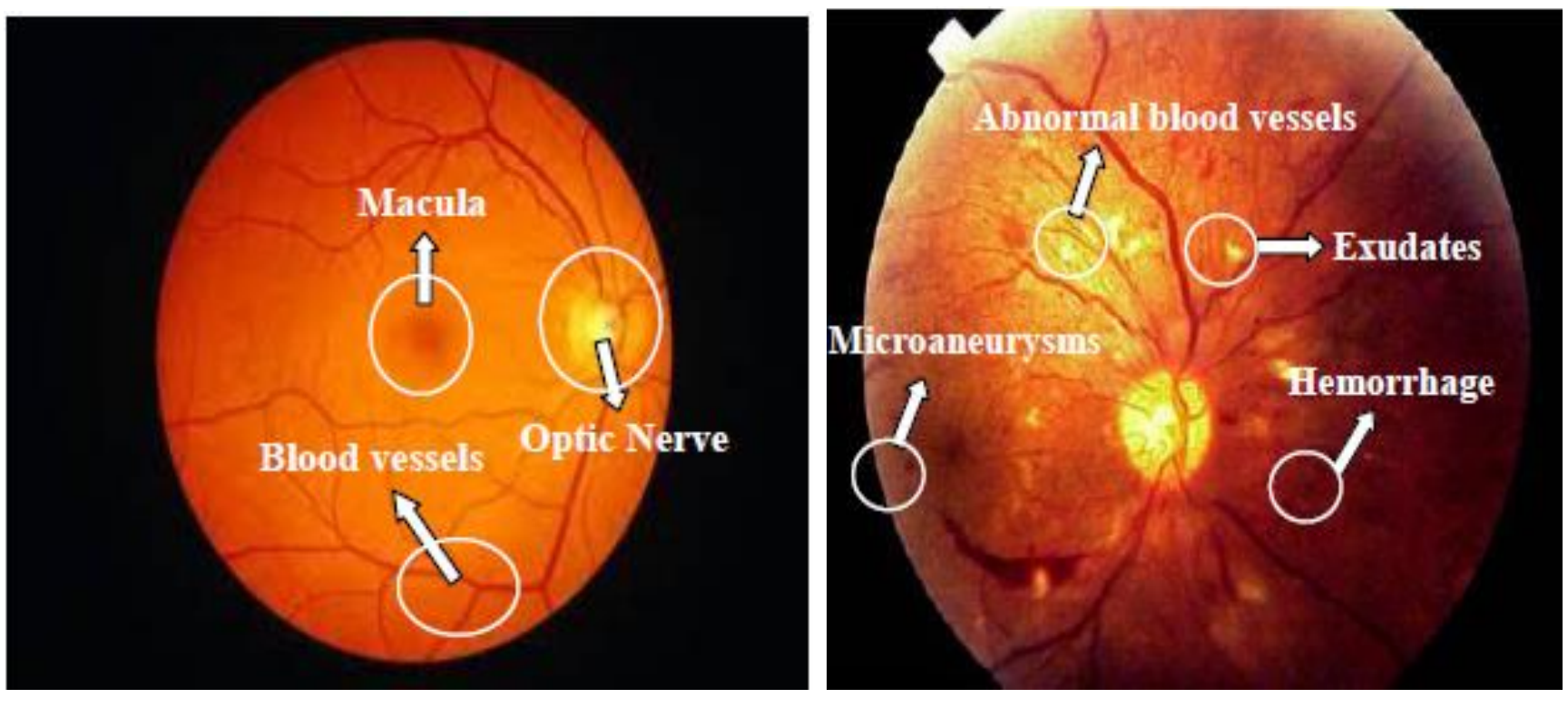

(a) (b)

Fig.1. Fundus Image (a) Normal; (b) DR-affected [10].

Based on structural differences in color fundus pictures, diabetic retinopathy is divided into two types: Non-proliferative Diabetic Retinopathy (NPDR) and Proliferative Diabetic Retinopathy (PDR). Hard exudates, Microaneurysms, Soft exudates, and Hemorrhages, are some of the symptoms of NPDR, while Neovascularization and Vitreous hemorrhage are indicators of PDR [10]. However, early identification of diabetic retinopathy is very important for preventing vision loss. For several reasons, such as a significant proportion of patients at a single institution or an essential and crucial patient condition, a quick clinical check and decision are frequently required. Screening a big group of people by hand is time-consuming and labor-intensive. Furthermore, all patients should receive affordable treatment; Patients in many underdeveloped nations do not have access to quality health care or expensive treatment. As a result of the absence of sufficient treatment, many indigent individuals are at risk of losing their sight. Consequently, there is a need for reliable auto DR screening methods using artificial intelligence to detect DR. In this work, we have developed an efficient model for detecting DR via an ensemble of different transfer learning models. It has contributed to an excellent outcome. The contribution of this article can be listed as follow as

- Proposing highly accurate transfer and ensemble model.
- Performance analysis of nine pre-trained model.
- Introduction regularization in each model.
- Replacement of fully connected layer with global average pooling layer
- Validating the average ensemble learning with weighted ensemble learning concept.
- Comparative analysis with the state of the existing work.

In the rest part of the article, Section 3 represents the Materials and Methodology. Section 4 shows the work results and discussion, and Section 5 contains the conclusion of the work.

## 2. Literature Review

Many deep learning-based research had been carried out on diabetic retinopathy (DR) detection from fundus images. This section discusses some of the existing research works. The likelihood of lesion patches was employed by Zago et al. [11] to categorize diabetic retinopathy or non-DR fundus pictures using two CNNs (pre-trained VGG16 and CNN). The DIARETDB1 dataset was utilized for training. The DDR, IDRiD, Messidor, DIARETDB0, Messidor-2, and Kaggle datasets were utilized for testing. The Messidor dataset delivered the best outcomes, with an AUC of 0.912 and a sensitivity of 0.94. A fundus image dataset can be classified as referable or non-referable DR using the model presented by Jiang et al. [12] using three CNNs (Inception-v3, ResNet152, and Inception-ResNet-v2). Prior to CNN training, the images were scaled, improved, and augmented, and the Adaboost approach was used to combine them. The network weights were updated using the Adam optimizer, and the system obtained an accuracy of 88.21 % and an AUC of 0.946. According to the DR severity levels, Kaggle fundus photos were divided into five classes by Pratt et al. [13] using CNN with 10 convolutional layers, 8 max-pooling layers, 3 fully connected layers, and a softmax classifier. Normalized and resized color fundus images L2 regularization and dropout methods were used to reduce overfitting. The model produced results with 95% specificity, 75% accuracy, and 30% sensitivity. Jayakumari.C et al. [14] proposed a transfer learning model where Inception V3 was used a pre-trained model and dropout layer was used to avoid overfitting. The model had a training accuracy of 98.6% using the Kaggle dataset. The model's accuracy for no DR is 86.6 %, mild is 62.5 %, moderate is 66.6 %, severe is 57.1 %, and PDR is 42.8 %. Shaohua Wan et al. [15] adopted AlexNet, VggNet, GoogleNet, ResNet with transfer learning and hyper-parameters tunning for analyzing diabetic image classification on Kaggle dataset. VggNet-s by hyper-parameters contributed best accuracy of 95.68. The severity of DR was classified by Narayana Bhagirath Thota et al. [16] using the VGG-16 model as a pre-trained neural network for fine-tuning. On high quality photos, data augmentation, batch normalization, dropout layers, and learn-rate scheduling were used to obtain an accuracy of 74%. Hasan Sabbir et al. [17] proposed an ensemble of SVM, KNN and Naïve Bayes model which was applied on MESSIDOR fundus dataset. It achieved 92% accuracy. A deep learning model incorporating transfer learning from VGG16 was created by Robiul Islam et al. [18]. With the new Kaggle dataset "APTOS 2019 Blindness Detection," it cut training time and produced average accuracy of 0.9132683. CNN (VGGnet) was utilized by Ahsan Habib Raj et al. [19] to estimate diabetic retinopathy (DR) and achieved 95.41% accuracy. Inception-ResNet-v2 was previously trained using transfer learning, then a custom block of CNN layers was built on top of Inception-ResNet-v2 to create the hybrid model, according to Kumar Gangwar et al. [20]'s proposal. On the Messidor-1 and APTOS datasets, the model has test accuracy of 72.33 % and 82.18 %, respectively. Sehrish Qummar et al. [21] trained an ensemble of five deep Convolution Neural Network (CNN) models (Resnet50, Xception, Inceptionv3, Dense169 and Dense121) using the publicly accessible Kaggle dataset of retina images and reached an accuracy of 80.70 %. In order to enhance image quality and consistently equalize intensities, Asra Momeni Pour et al. [22] created a new diabetic retinopathy monitoring model that used the Contrast Limited Adaptive Histogram Equalization approach. The EffcientNet-B5 architecture is then used for the classification step. This network's effectiveness lies in its ability to scale all of its dimensions consistently. The final model is initially trained using a blend of the Messidor-2 and IDRiD datasets, and then it is validated on the Messidor dataset. The area under the curve (AUC) is raised to 0.945 from 0.936, the maximum value in all recent works. Convolutional Block Attention Module (CBAM) was built on top of the encoder by Mohamed M. Farag et al. [23] to increase its discriminative power. They used the encoder from DenseNet169 to generate a visual embedding. Applying on APTOS dataset, the model contributed 97% accuracy. A summary of review work is presented in the Table 1.

TABLE 1: Summary of related work

| Ref. | Model | Dataset | No. of Image | Augmentation | Transfer Learning | Ensemble Learning | Accuracy (%) |
|---|---|---|---|---|---|---|---|
| [17] | SVM+ KNN+Naïve Bayes | Messidor dataset | 1200 | No | No | yes | 92 |
| [18] | VGG16 | APTOS 2019 Blindness Detection | 5590 | No | yes | No | 80 |
| [19] | VGGnet | Kaggle | 35126 | No | No | No | 95.41 |
| [24] | VGG-NiN | EyePACS | 88,702 | No | No | No | 85 |
| [20] | Inception-ResNet-v2 | Messidor-1<br>APTOS | 1200<br>5590 | yes | yes | No | 72.33<br>82.81 |
| [21] | Resnet50, Inceptionv3, Xception, DenseNet121, DenseNet169 | Kaggle | 35126 | yes | No | Yes | 80.7 |
| [22] | EfficientNet-B5 | Mixture of Messidor-2 and IDRiD | 1748<br>516 | No | No | No | 94.5 |
| [23] | DenseNet169 and Convolutional Block Attention Module | APTOS | 5590 | No | yes | No | 97 |

From the literature, it is obvious that initially the researcher used traditional ML methods in DR detection. Day by day, CNN, transfer learning approach were being popular. Regularization, replacement of fully connected layer by global average pooling layer, updated pre-trained models and ensemble learning concept are not used in diabetic retinopathy detection. The performance of those study was also not so high. In this study, we have included and resolved the issues and a comprehensive analysis has been carried out.

## 3. Materials and Methodology

In order to diagnose diabetic retinopathy from fundus images, this research offers nine transfer learning models. Combination of two publicly available dataset has been used to carry out these experiments. The following section describes the whole methods and experimental setup in details.

### 3.1 Dataset

For this experiment, we combined the Diseases Grading of Indian Diabetic Retinopathy Image Dataset (IDRID) [25] with the fundus-dataset from Mendeley [26]. 1500 genuine color fundus photos in 24-bit RGB format are divided into 26 categories in the dataset [26]. Diseases Grading of IDRID consists of original color fundus images (516 images divided into train set (413 images) and test set (103 images) [25]. We have taken 300 normal fundus images from Mendeley dataset and trainset of 431 diabetics fundus images from IDRID dataset. So, our dataset consists of two class named “Normal” and “Diabetic”. Fig.2 displays the distribution of images by ease class. It is obvious that our dataset is quite small compared to the Messidor dataset [27]. Whatever, the main aim of our work is to design robust and accurate model with the limited label data.

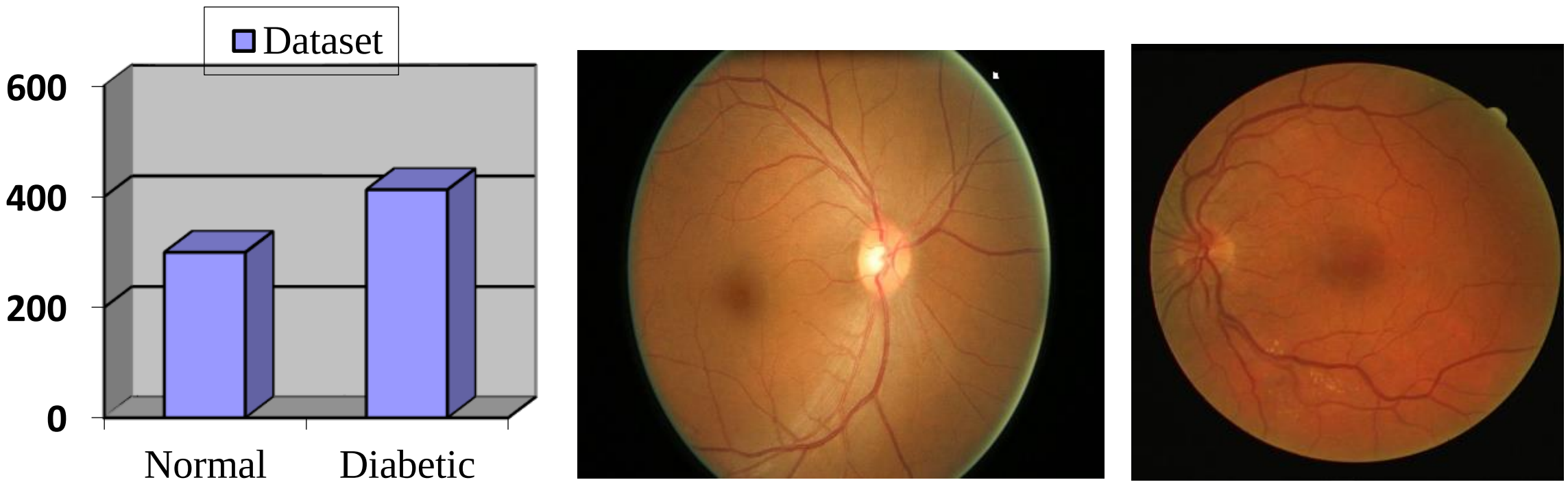

(a) (b) (c)

Fig.2. (a) Distribution of Data in each class; (b) Normal Image of Fundus; (c) Diabetic Image of Fundus.

### 3.2 Data Augmentation

On numerous Computer Vision projects, Deep Convolutional Neural Networks (DCNNs) have displayed impressive performance. However, these systems are intensely dependent on enormous dataset to dodge over fitting [28]. When a network learns a function with significant variance, the behavior is known as overfitting. It occurs due to lack of training dataset and insufficient diversity of training data or uneven class balance in dataset [29]. augmentation is a way to deal such problem. The performance of DCNN will be enhanced with the increasing of training dataset with different augmentation technique [30]. The diabetic retinopathy dataset was split into training, validation and testing data folder using splitfolders.ratio() module in Python at ratio 0.70, 0.15 and 0.15 respectively and resized at 224*224 pixels. Table 2. contains the summary of datasets. Data augmentation techniques such as rotation, zooming, flipping, shifting was applied in training dataset in our work by using ImageDatagenerator module of python [31]. The parameters of augmentation are listed in Table 3.

TABLE 2: Training, Validation and testing data distribution.

| SI. No. | Dataset | No of Images | | |
|---|---|---|---|---|
| | | Normal | Diabetic | Total |
| 1 | Training | 210 | 289 | 499 |
| 2 | Validation | 45 | 61 | 106 |
| 3 | Testing | 45 | 63 | 108 |

TABLE 3: Data augmentation Techniques with parameters.

| SI. No. | Augmentation Techniques | Parameters |
|---|---|---|
| 1 | Rotation | rotation_range = 10 |
| 2 | Zoom | zoom_range = 0.2 |
| 3 | Width shift | width_shift_range = 0.2 |
| 4 | Height shift | height_shift_range = 0.2 |
| 5 | Vertical flip | vertical_flip = True |
| 6 | Horizontal flip | horizontal_flip = True |

### 3.3 Transfer Learning

Using a learning strategy created for one assignment as the basis for a model on another assignment is called transfer learning (TL), a machine learning methodology [32]. It reuses the pre-trained model on new problem. The main benefits of TL are reduced training time, improved neural network performance (in most cases), and not needing of a large amount of data [33]. The most common pre-trained models for TL are VGG19, VGG16, AlexNet and Inceptions etc. [32]. In this work, ResNet50, InceptionV3, Xception, MobileNetV2, NASNetMobile, VGG19, DenseNet121, DenseNet169 and DenseNet201 were applied as pre-trained model. The classification layer of each model is replaced by GlobalAveragePooling2D layer, SoftMax layer, Dense layer with the number of 2 classes and Dropout layer. Dropout value 0.25 was carried out due to avoid overfitting. All the models were fitted using Adam optimizer with learning rate of 0.001, categorical cross-entropy, and batch size of 16. The table 4 shows the trainable parameters of the models. Among all the pre-trained models, DenseNet121 provided the best outcome with 100% accuracy and DenseNet169 and DenseNet201 performed also pretty better and have chosen for ensemble learning.

TABLE 4: TL architectures applied in our works

| Model | Total Parameter | Trainable Parameters | Non-Trainable Parameters |
|---|---|---|---|
| InceptionV3 | 21,806,882 | 4,098 | 21,802,784 |
| Xception | 20,865,578 | 4,098 | 20,861,480 |
| DenseNet121 | 18,321,984 | 2,050 | 7,037,504 |
| DenseNet201 | 18,325,826 | 3,842 | 18,321,984 |
| DenseNet169 | 12,646,210 | 3,330 | 12,642,880 |
| ResNet50 | 23,591,810 | 4,098 | 23,587,712 |
| NASNetMobile | 4,271,830 | 2,114 | 4,269,716 |
| VGG19 | 20,025,410 | 1,026 | 20,024,384 |
| MobileNetV2 | 2,260,546 | 2,562 | 2,257,984 |

### 3.4 Ensemble Learning

Ensemble learning attempts to outperform any single algorithm by integrating several algorithms and combining the results with various voting processes [34]. It is used to decrease variance, bias and improve the prediction. [21]. From the performance Table 5 and confusion matrix analysis, TL DenseNet169 and DenseNet201 architectures made prediction very well. So, Average ensemble and weighted ensemble both were performed on the models separately. Average ensemble and weighted ensemble are shown in algorithm 1 and 2.

**Algorithm 1. Average Ensemble of the Models**

**Input:** Test_set $T$: Models $Z_j$ (j $=$ 1 to m) where j is the number of models

**Output:** $I_o$

Ensemble_model B= [$Z_1$,$Z_2$, . . .$Z_j$ ]

**For** n = 1 to k **do**

  Predict, $P$ = generate ($T$)

  $Q$= add ($P$, along y axis)

  $I_o$= index_max (Q, along x axis)

Confusion_matrix ($I_o$, $T$)

Classification_ matrices ($I_o$, $T$)

**End**

**Algorithm 2. Weighted ensemble of the models**

**Input:** Test_set $T$: Models $Z_j$ and Weight_set $W_j$ (j $=$ 1 to m) where j is the number of models.

**Output:** $I_o$

Ensemble_model B= [$Z_1$,$Z_2$, . . .$Z_j$ ]

**For** n = 1 to k **do**

  Predict, $P$ = generate ($T$)

  $Q = add\ (P * W_i, along\ y\ axis)$

  $I_o$= index_max (Q, along x axis)

Confusion_matrix ($I_o$, $T$)

Classification_ matrices ($I_o$, $T$)

**End**

In our work, E= [DenseNet169, DenseNet201] and Weight $W_k = [\ 0.2\ ,02]$. K=1,2.

### 3.5 Proposed Methodology

This work represents a TL model and an Ensemble model in the detection of diabetic retinopathy shown in Fig.3. IDRID dataset from Kaggle and Fundus-Dataset from Mendeley were combined and split into training, validation and testing with ratio of 70%, 15%, 15% respectively. The training dataset was augmented with parameter tuning using the techniques described in previous section. The nine pre-trained models in Table 4. were carried out in this experiment adding Average Global Pooling Layer, Dropout layer & dense layer at bottom of base models. The Adam optimizer and categorical cross-entropy loss function were used to precisely train the networks. The categorical entropy loss can be expressed by the following equation (1).

$$loss = -\sum_{i=1}^{Output\ size} y_i * log_{\hat{y}_i} \qquad (1)$$

Where $\hat{y}_i$is the $i-th$ scalar value in the model output, $y_i$ is the corresponding target value, and output size is the number of scalar values in the model output.

The models were trained, validated and tested. Out of them, the best model was found out. Comparatively two little poor model were ensembled to generate new classifier. All the classifiers were evaluated using performance measures such as a confusion matrix, precision, recall, F1-score, and accuracy to choose the optimal model for Diabetic retinopathy diagnosis.

### 3.6 Performance Evaluation

Our proposed model's performance was evaluated both qualitatively and visually. The qualitative evaluation of image classification is a widely used method [35]. In addition, we quantified our model by measuring the parameters Accuracy (Acc), Precision, recall, and F1-score.

$$\text{Accuracy} = \frac{\sum TP + \sum TN}{\sum TP + \sum TN + \sum FP + \sum FN} * 100 \qquad (2)$$

$$\text{Precision} = \frac{\sum TP}{\sum TP + \sum FP} * 100 \qquad (3)$$

$$\text{Recall} = \frac{\sum TP}{\sum TP + \sum FN} \qquad (4)$$

$$\text{F1 score} = 2 * \left(\frac{Precision * Recall}{Precision + Recall}\right) * 100 \qquad (5)$$

$$\text{Sensitivity} = \frac{TP}{TP + FN} \qquad (6)$$

$$\text{Specificity} = \frac{TN}{FP + TN} \qquad (7)$$

$$\text{Macro Avg Measure} = \frac{1}{N}(Measure\ in\ class_1 + Measure\ in\ class_2 + \cdots + Mesure\ in\ class_N) \qquad (8)$$

$$\text{Weighted Average Measure} = \frac{1}{Total\ number\ of\ sample}\ [(Measure * weight)\ in\ Class_1 + (Measure * weight) in\ class_2 + \cdots + (Measure * weight) in\ class_N] \qquad (9)$$

Where: TP stands for True Positive, TN denotes True Negative, FP is False positive and FN is False Negative.

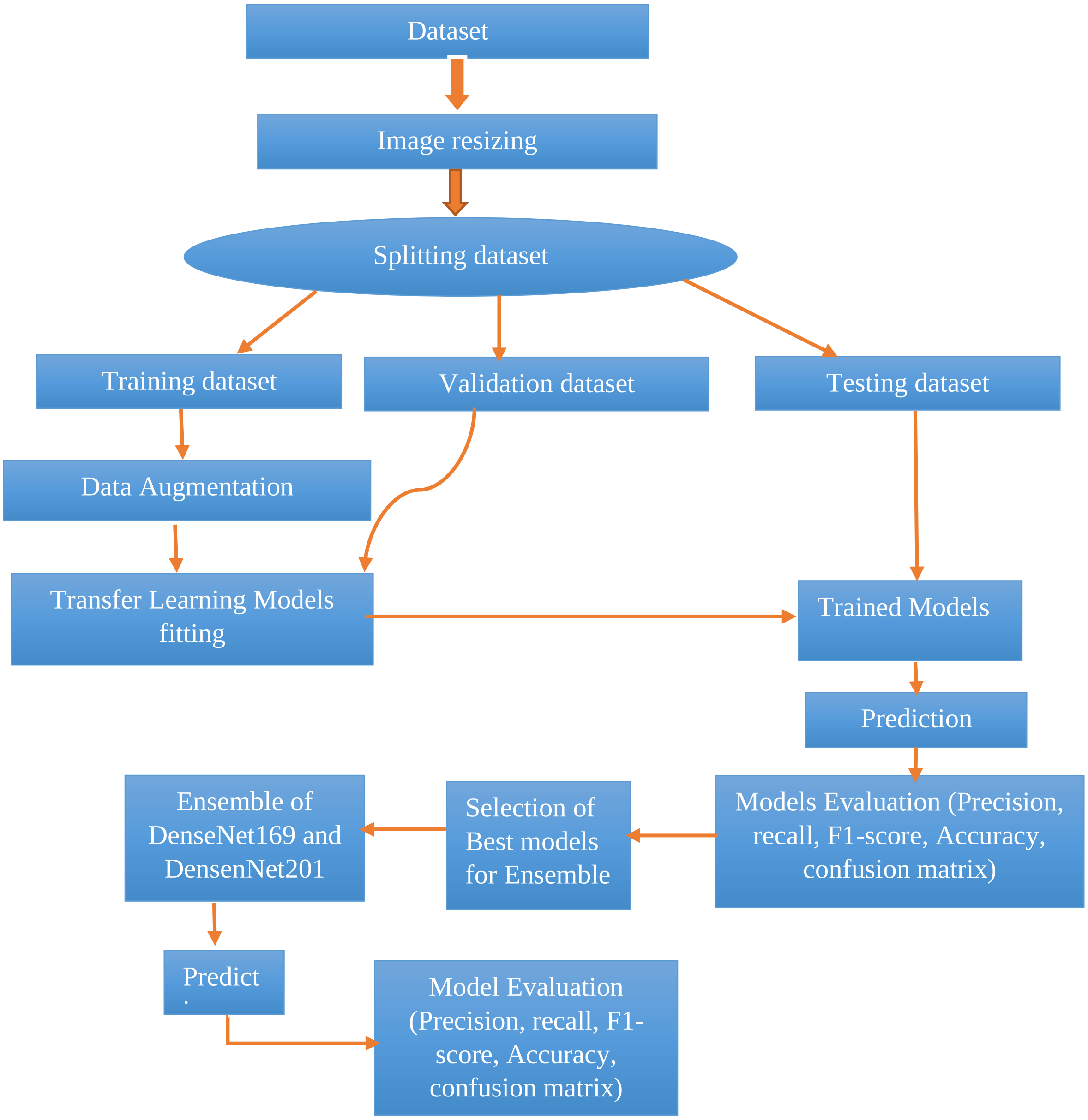

Fig.3. Proposed Methodology.

## 4 Result and Discussion

In this work, Different transfer learning architectures are implemented in the detection of diabetic retinopathy. It was carried out on GoogleColab with GPU. Different pre- trained models such as DesnseNet121, DenseNet201, DenseNet169, InceptionV3, Xception, ResNet50, NASNetMobile, VGG19 and MobileNetV2 were used with adding Global Average Pooling layer, dropout layer and Dense layer at the beneath of base model. Data Augmentation with hyper parameters tuning in Table 3 was applied on training dataset. The Training loss and validation loss are presented in the Fig. 4. In the case of inceptionV3, Xecption, ResNet50, Validation loss is smaller than training loss. Overfitting exists here and Underfitting is in VGG19 since validation loss is quite greater than training loss. The rest are reached almost at optimal fit. Among them DenseNet201, DenseNet169, DenseNet121 are excellent and DenseNet121 is the best in optimal model fit. Similarly in Fig. 5. In case of InceptionV3, Xecption, ResNet50 and VGG19, the model fit in training and validation accuracy is quite poor. The rest are pretty good but the DenseNet121, DenseNet201 and DenseNet169 are the superior. Fig.6. represents the confusion matrix for each TL model and Ensemble Learning model where the number of classification and misclassification are easily visible. ResNet50 shows the highest misclassification number of 22. DenseNet121 provides the best prediction where all the testing data has classified properly. Average ensemble and weighted ensemble with same weight of DenseNet201 and DenseNet169 also result the exact classification. Precision, recall, f1 score and accuracy has been calculated for individual class in Table 5. DenseNet121, DenseNet169, DenseNet201, InceptionV3 NASNetMobile, VGG19 and Ensemble models show the 100% precision but NASNetMobile, VGG19 show lower recall and F1-score on 'Normal' class dataset. DesneNet121, ensemble models, Xception and MobileNetV2 outperform the rest in precision but Xception and MobileNetV2 result the lower recall and F1 score. From the analysis, DenseNet121 and Ensemble models provide 100% precision, recall, F1-score on both class in Table 5 and also remain same in overall performance in Table 6. So DenseNet121 and Ensemble models are taken as benchmark for DR detection.

Table 5: Performance Analysis of each model on each class.

| **Base Model** | **Normal** | | | **Diabetic** | | | **Accuracy (%)** |
|---|---|---|---|---|---|---|---|
| | **Precision** | **Recall** | **F1- score** | **Precision** | **Recall** | **F1- score** | |
| Dense Net169 (TL) | 100 | 98 | 99 | 98 | 100 | 99 | 99.07 |
| DenseNet201 (TL) | 100 | 98 | 99 | 98 | 100 | 99 | 99.07 |
| InceptionV3 | 100 | 98 | 99 | 98 | 100 | 99 | 99.07 |
| Xception | 98 | 100 | 99 | 100 | 98 | 99 | 99.07 |
| ResNet50 | 91 | 69 | 79 | 71 | 92 | 80 | 79.62 |
| NASNetMobile | 100 | 96 | 98 | 97 | 100 | 98 | 98.14 |
| VGG19 | 100 | 92 | 96 | 94 | 100 | 97 | 96.29 |
| MobileNetV2 | 96 | 100 | 98 | 100 | 97 | 98 | 98.14 |
| **DenseNet121 (TL)** | **100** | **100** | **100** | **100** | **100** | **100** | **100** |
| **Average Ensemble of DenseNet169 (TL) and DeseNet201 (TL)** | **100** | **100** | **100** | **100** | **100** | **100** | **100** |
| **Weighted Ensemble of DenseNet169 (TL) and DeseNet201 9 (TL)** | **100** | **100** | **100** | **100** | **100** | **100** | **100** |

Table 6: Overall performance Analysis of each model.

| **Base Model** | **Macro average** | | | **Weighted average** | | | **Accuracy (%)** |
|---|---|---|---|---|---|---|---|
| | **Precision** | **Recall** | **F1-score** | **Precision** | **Recall** | **F1-score** | |
| DenseNet169(TL) | 99 | 99 | 99 | 99 | 99 | 99 | 99 |
| DenseNet201(TL) | 99 | 99 | 99 | 99 | 99 | 99 | 99 |
| InceptionV3 | 99 | 99 | 99 | 99 | 99 | 99 | 99 |
| Xception | 99 | 99 | 99 | 99 | 99 | 99 | 99 |
| ResNet50 | 81 | 81 | 80 | 82 | 80 | 80 | 80 |
| NASNetMobile | 98 | 98 | 98 | 98 | 98 | 98 | 98 |
| VGG19 | 97 | 96 | 96 | 97 | 96 | 96 | 96 |
| MobileNetV2 | 98 | 98 | 98 | 98 | 98 | 98 | 98 |
| **DenseNet121** | **100** | **100** | **100** | **100** | **100** | **100** | **100** |
| **Average Ensemble of DenseNet169 (TL) and DeseNet201 (TL)** | **100** | **100** | **100** | **100** | **100** | **100** | **100** |
| **Weighted Ensemble of DenseNet169 (TL) and DeseNet201 9 (TL)** | **100** | **100** | **100** | **100** | **100** | **100** | **100** |

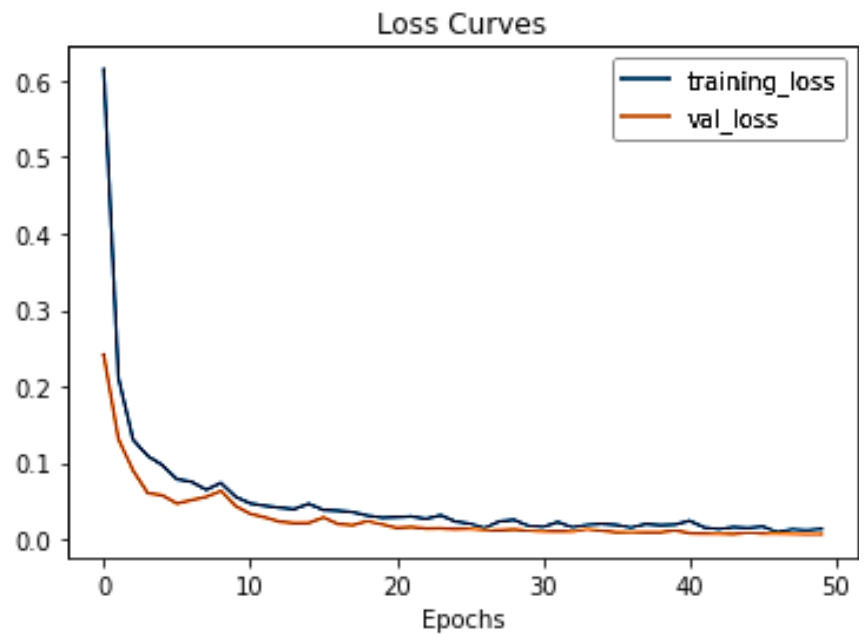

(a) DenseNet169

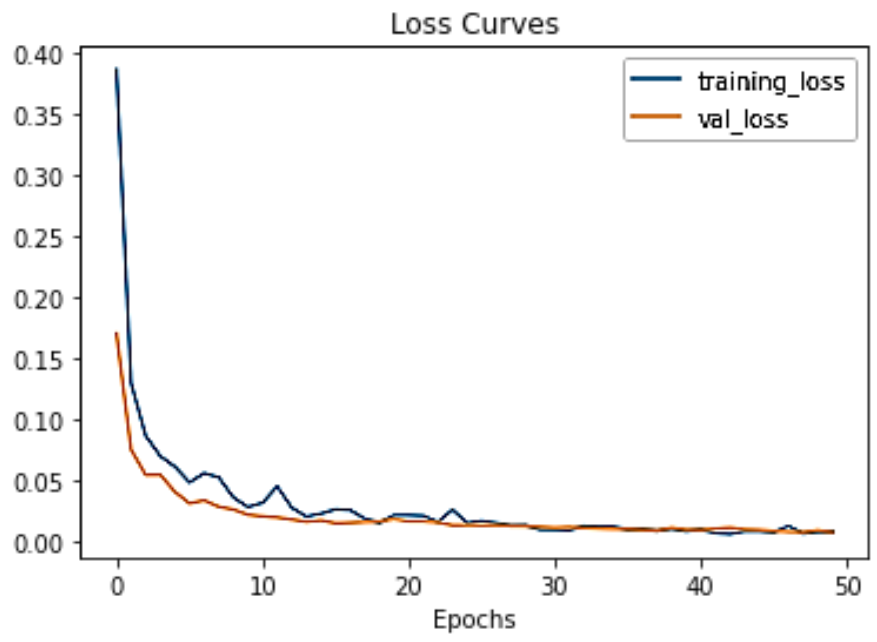

(b) DenseNet201

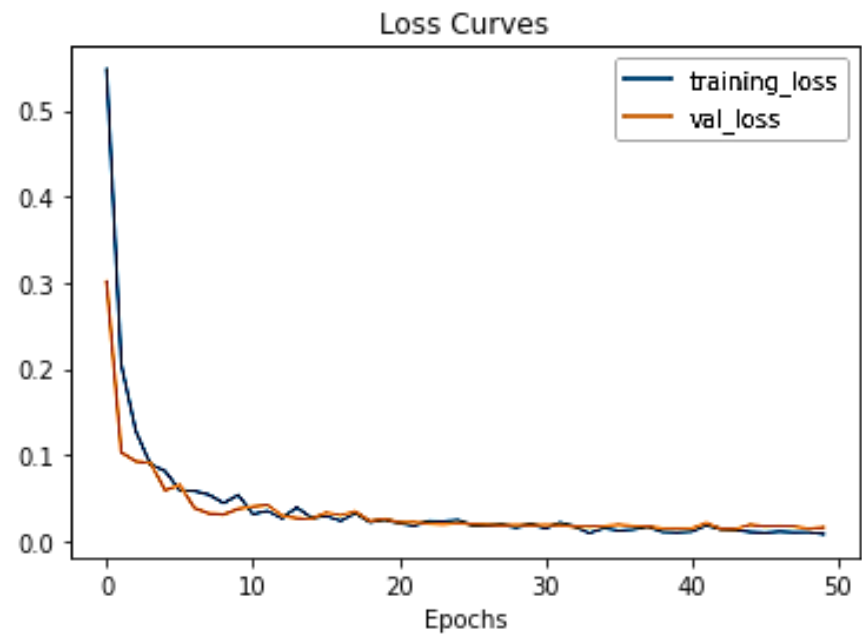

(c) DenseNet121

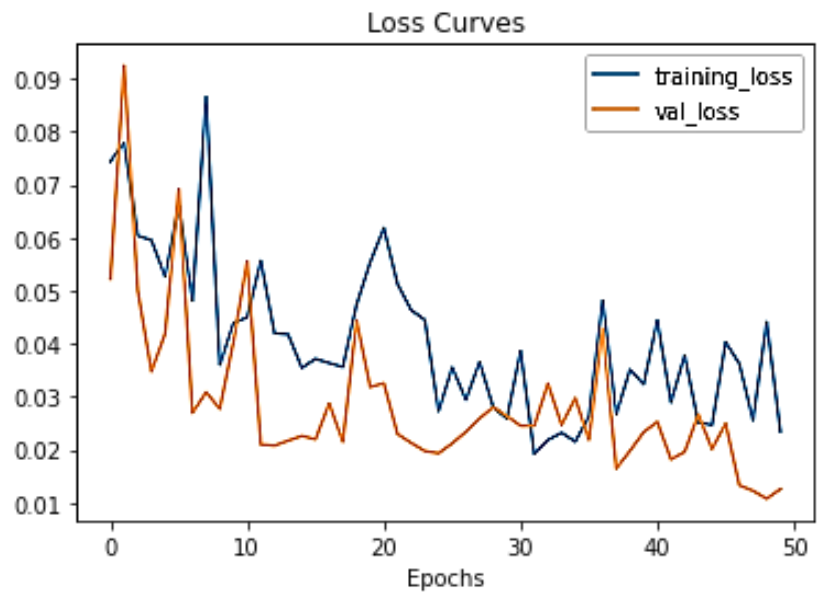

(d) InceptionV3

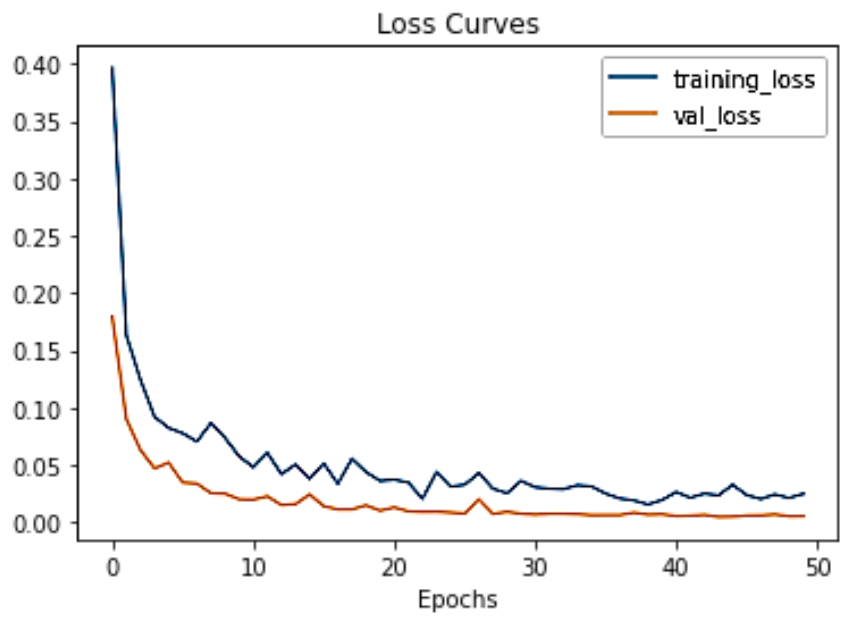

(e) Xception

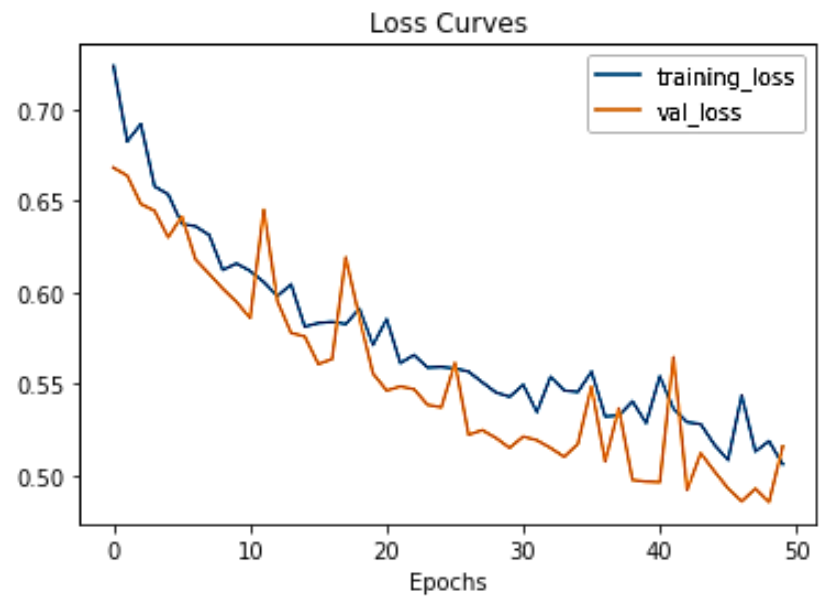

(f) ResNet50

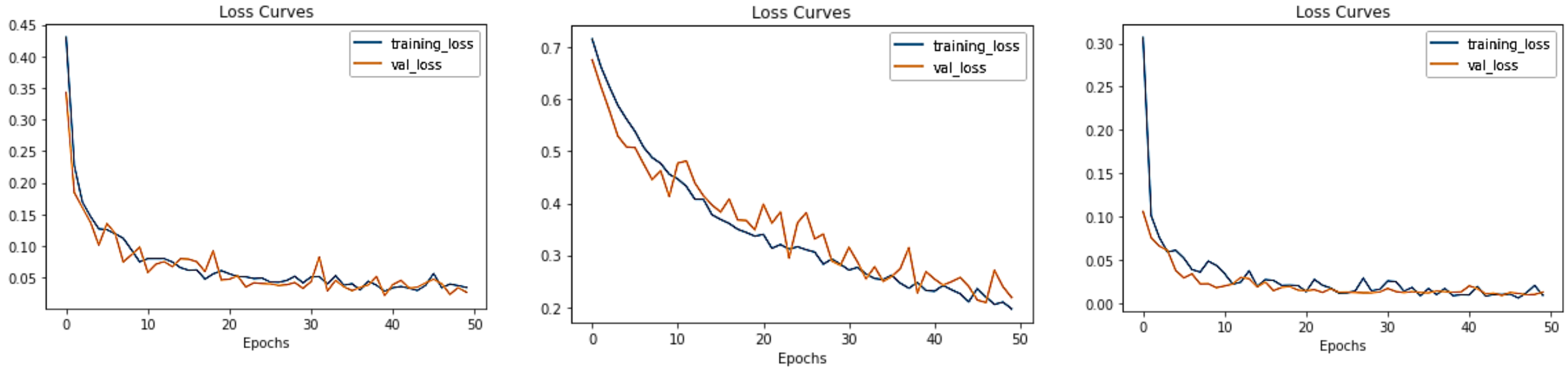

(g) NASNetMobile (h) VGG19 (i) MobileNetV2

Fig. 4. Training loss and Validation Loss when base model (a) DenseNet169; (b)DenseNet201; (c) DenseNet121; (d) InceptionV3; (e) Xception; (f) ResNet50; (g) NASNetMobile; (h) VGG19; (i) MobileNetV2

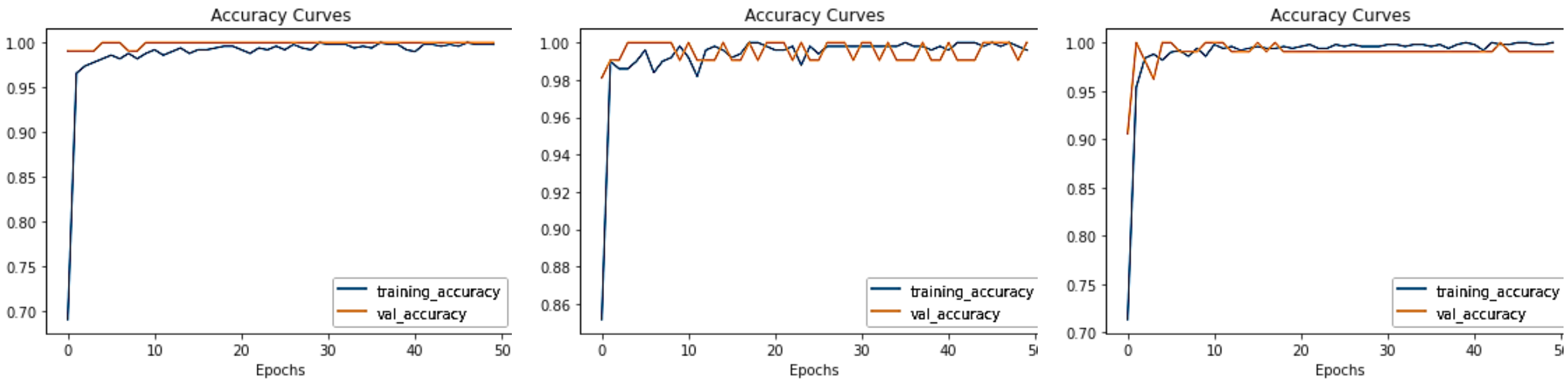

(a) DenseNet169 (b) DenseNet201 (c) DenseNet121

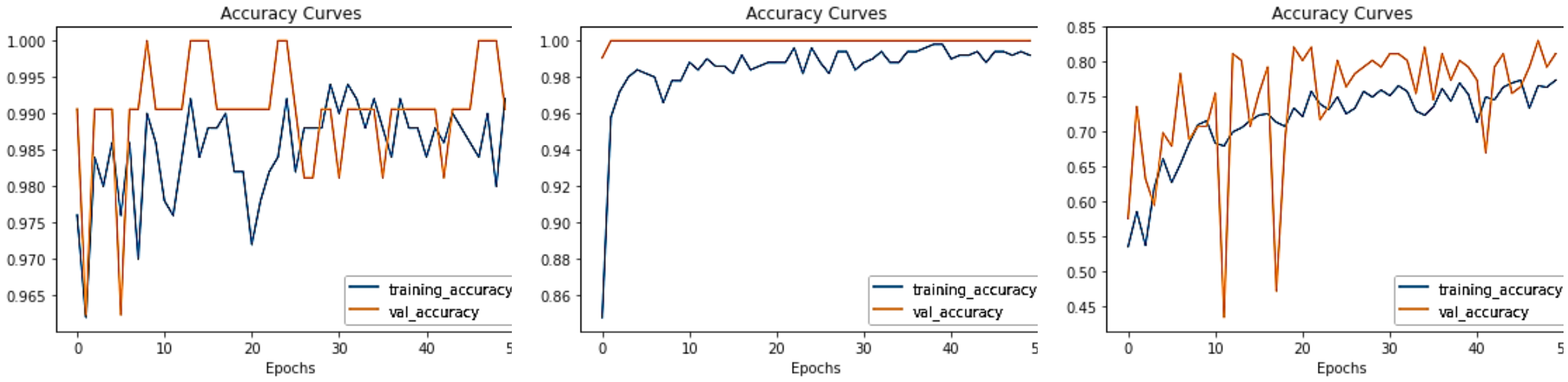

(d) InceptionV3 (e) Xception (f) ResNet50

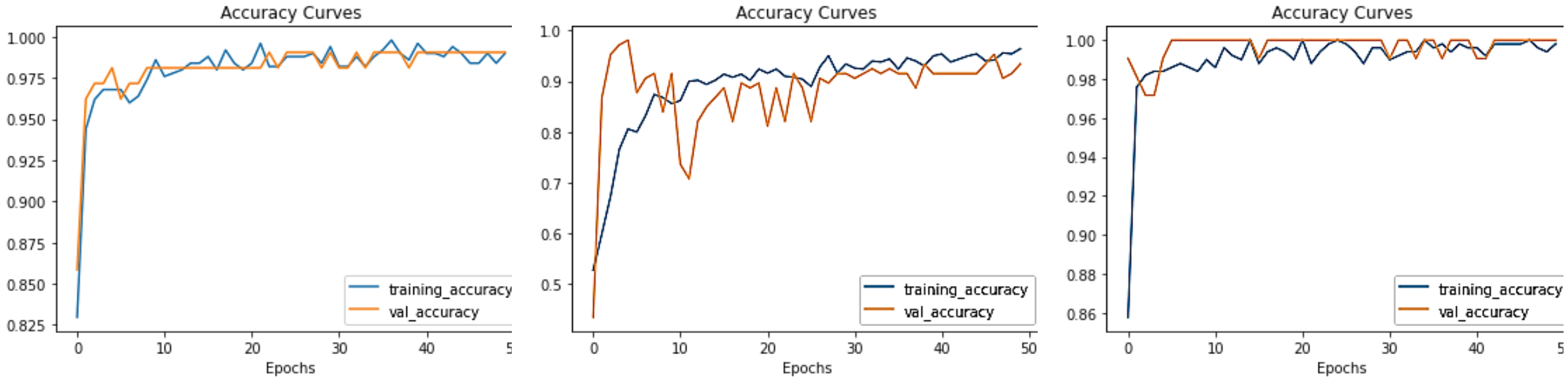

(g) NASNetMobile (e) VGG19 (i) MobileMetV2

Fig. 5. Training accuracy and Validation accuracy when base model (a) DenseNet169; (b)DenseNet201; (c) DenseNet121; (d) InceptionV3; (e) Xception; (f) ResNet50; (g) NASNetMobile; (h) VGG19; (i) MobileNetV2

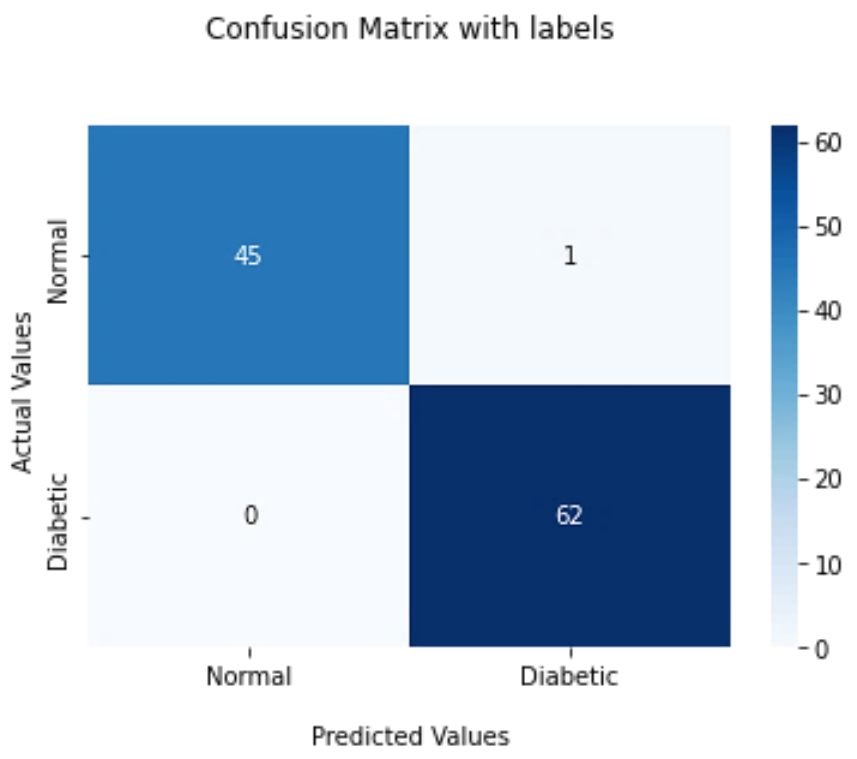

(a) DenseNet169

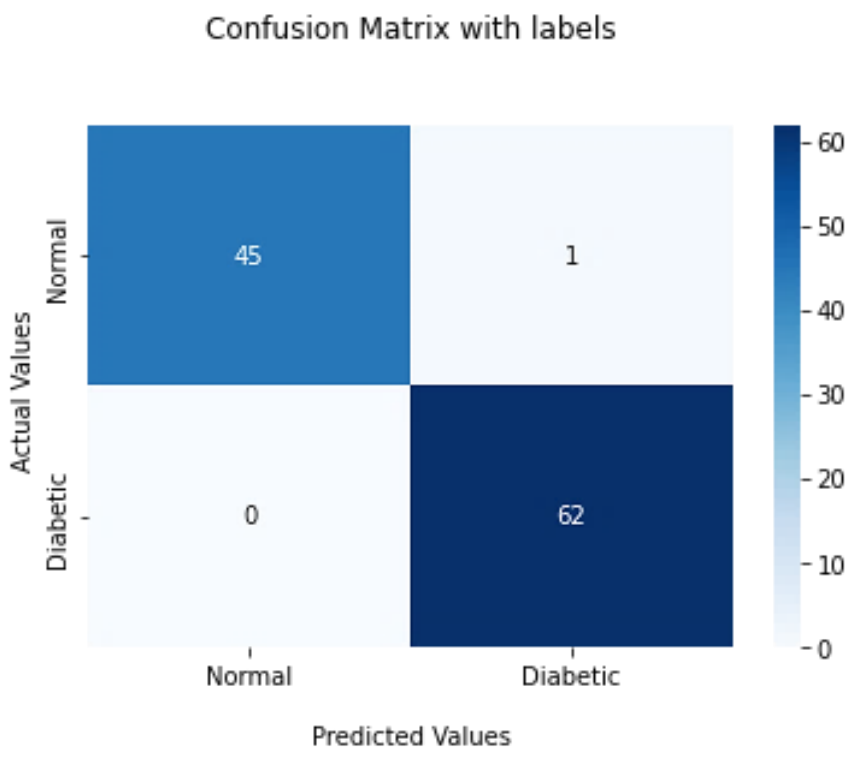

(b) DenseNet 201

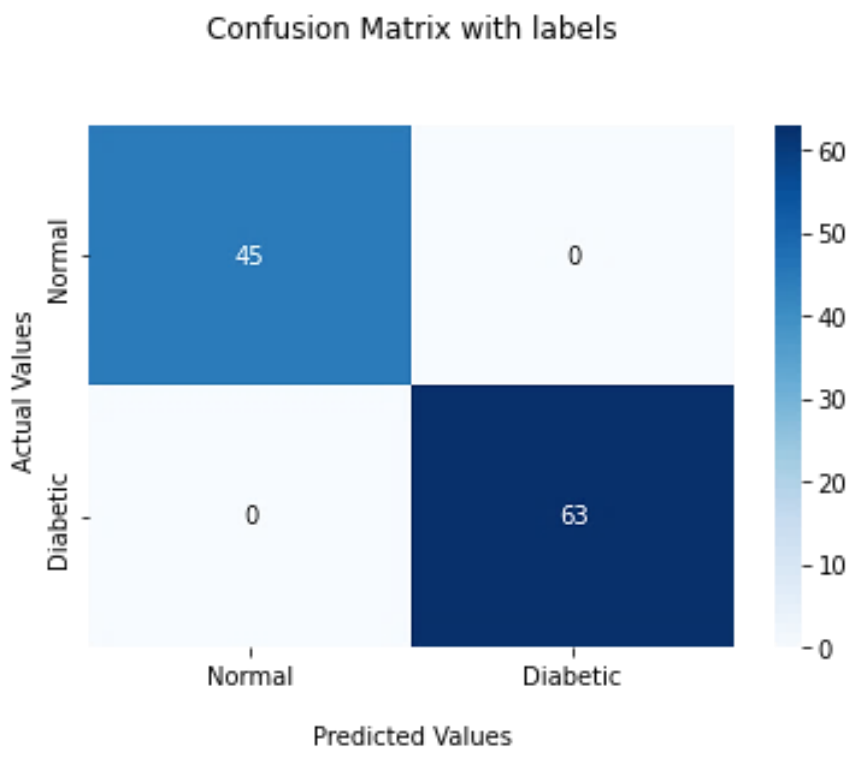

(c) Average Ensemble of DenseNet169 and 201

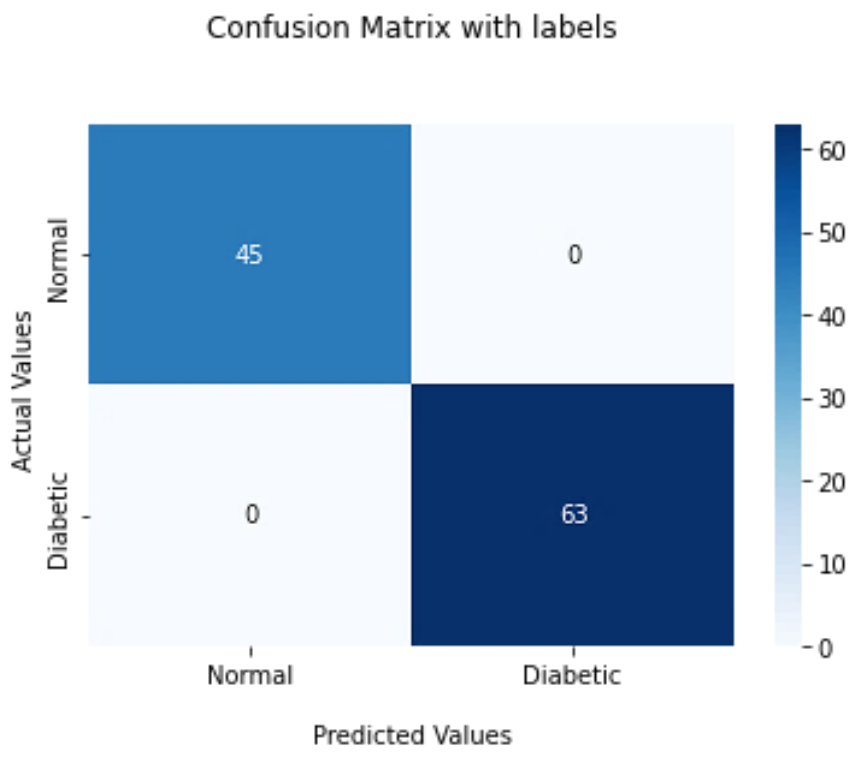

(d) Weighted Ensemble of DenseNet169 and 201

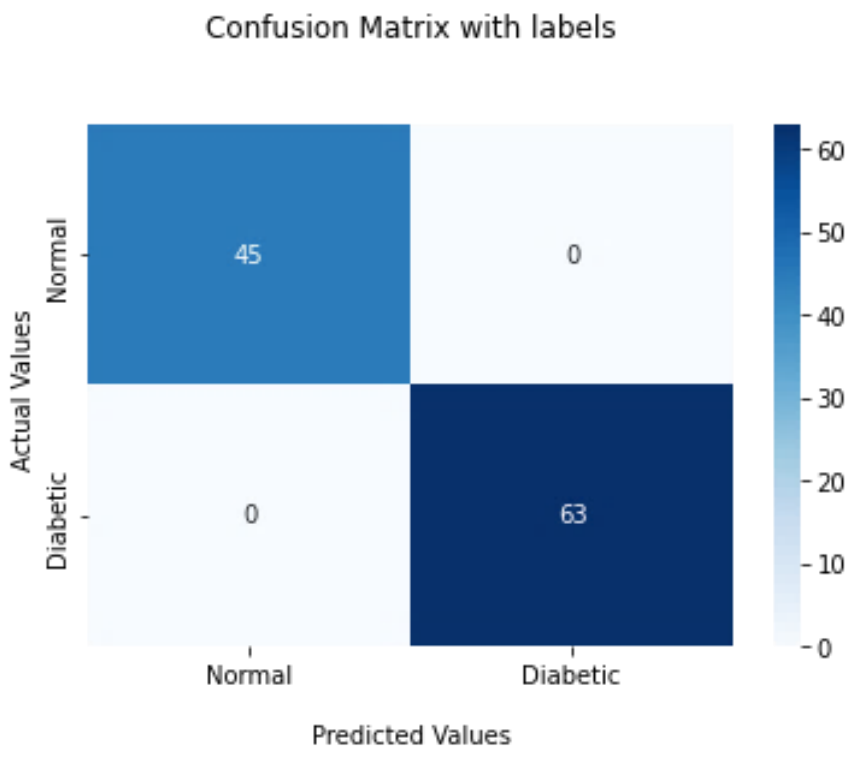

(e) DenseNet121

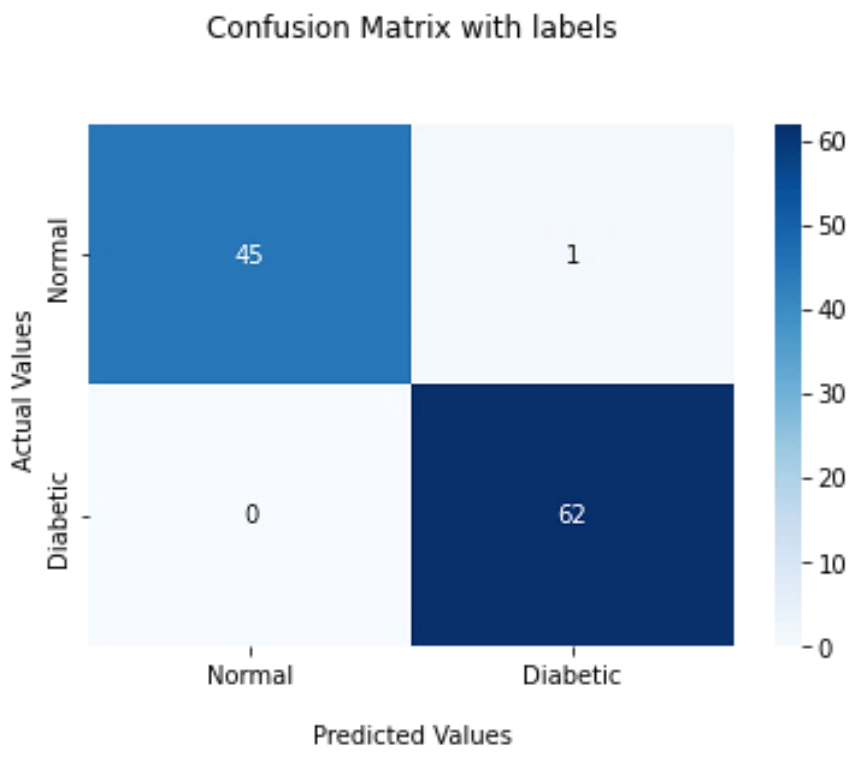

(f) InceptionV3

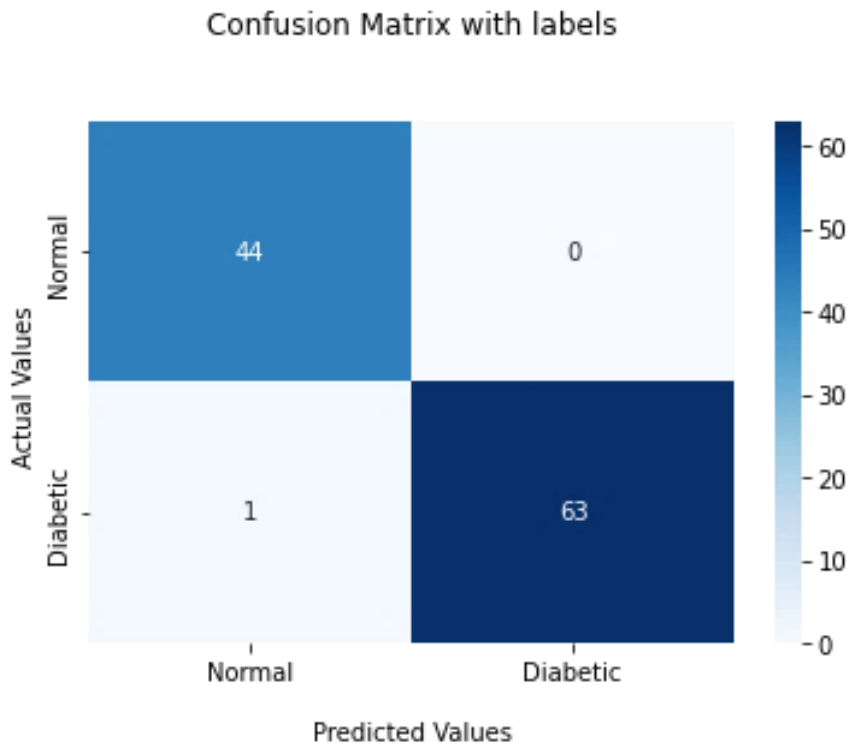

(g) Xception

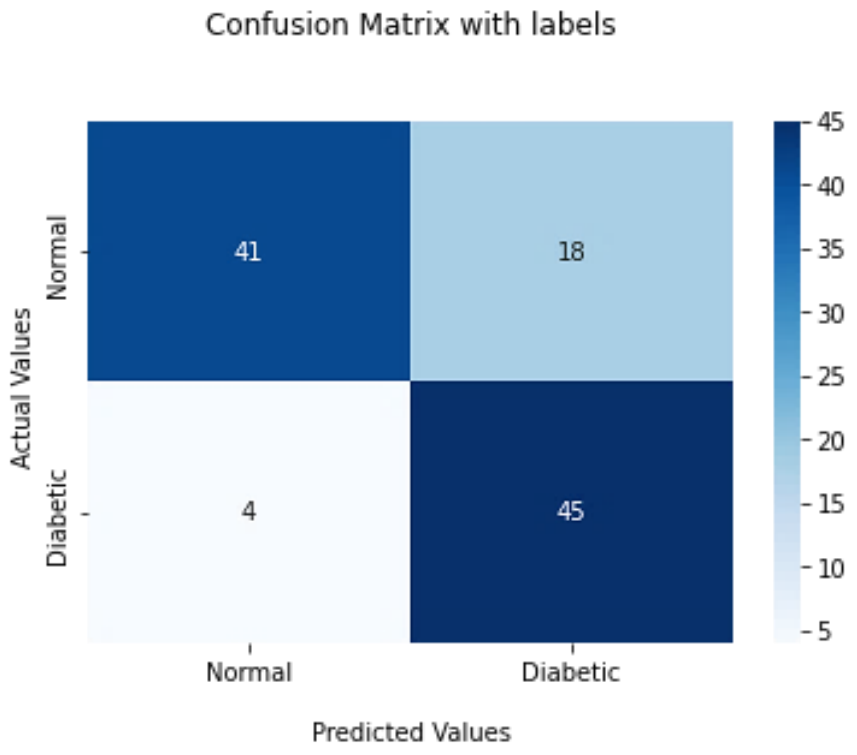

(h) ResNet50

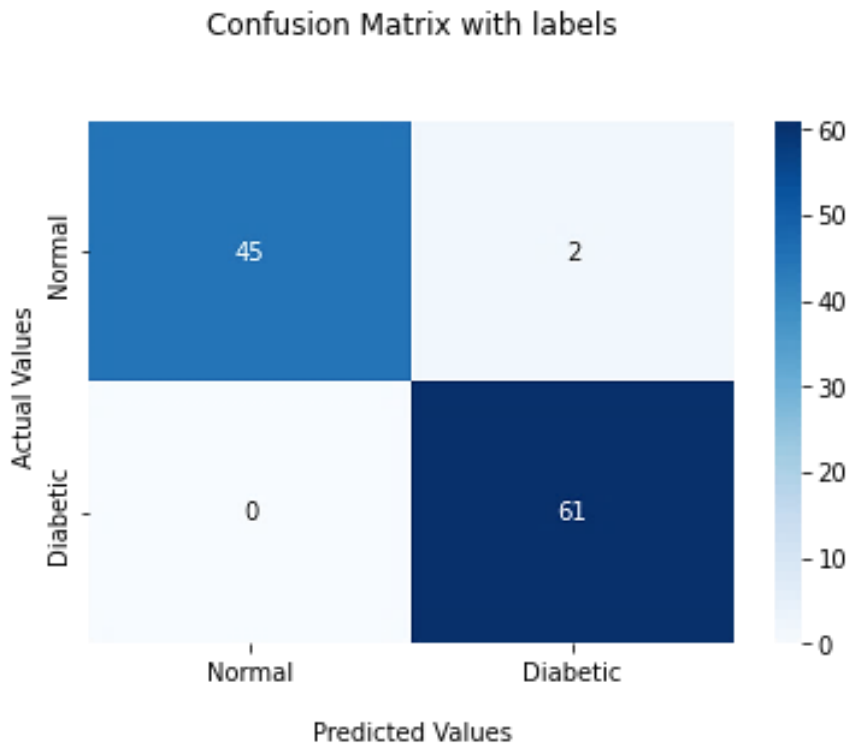

(i) NASNetMobile

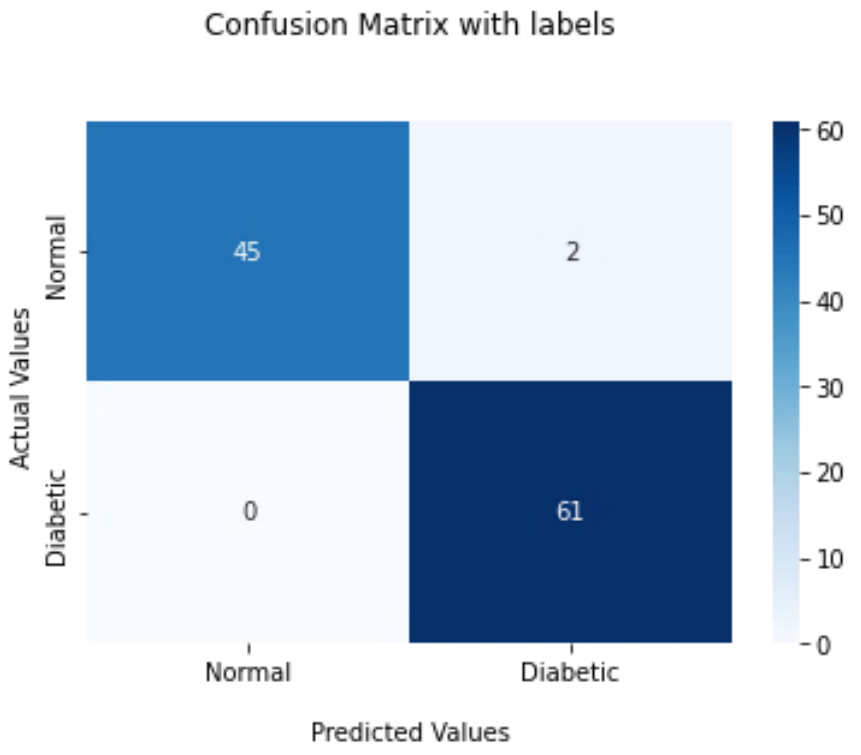

(j) VGG19

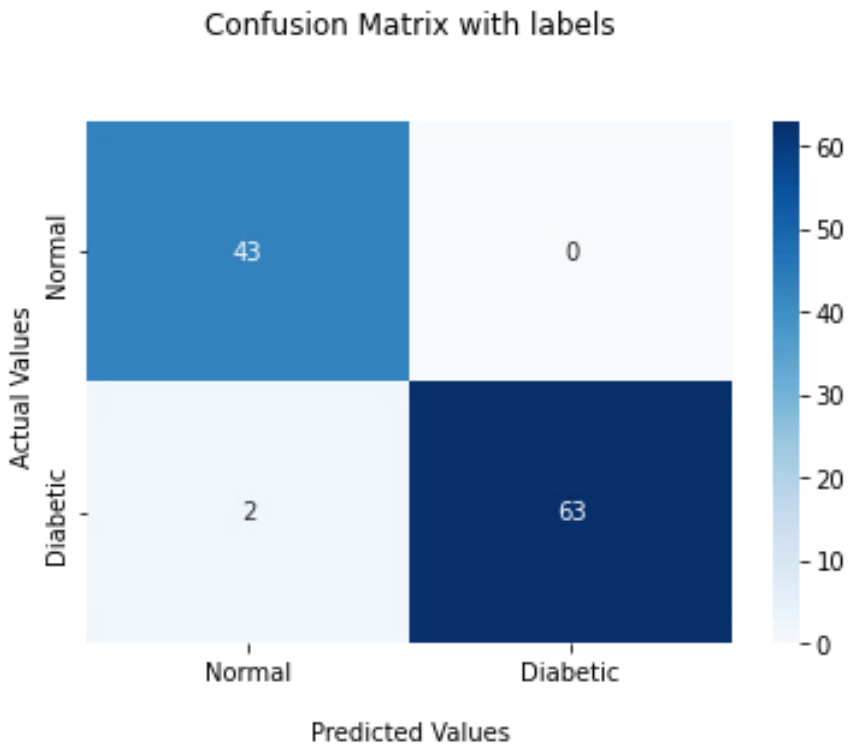

(k) MobileNetV2

Figure 6. Confusion Matrix. (a) DenseNet169; (b) DenseNet201; (c) Average Ensemble of DenseNet169 and 201; (d) Weighted Ensemble of DenseNet169 and 201; (e) DenseNet121; (f) InceptionV3; (g) Xception; (h) ResNet50; (i) NASNetMobile; (j) VGG19; (k) MobileNetV2

Fig.7. represents the comparison of our models in testing accuracy. Obviously, our proposed DenseNet169 TL learning model outperforms the other models. A comparison with the existing work is also shown in Table 7.

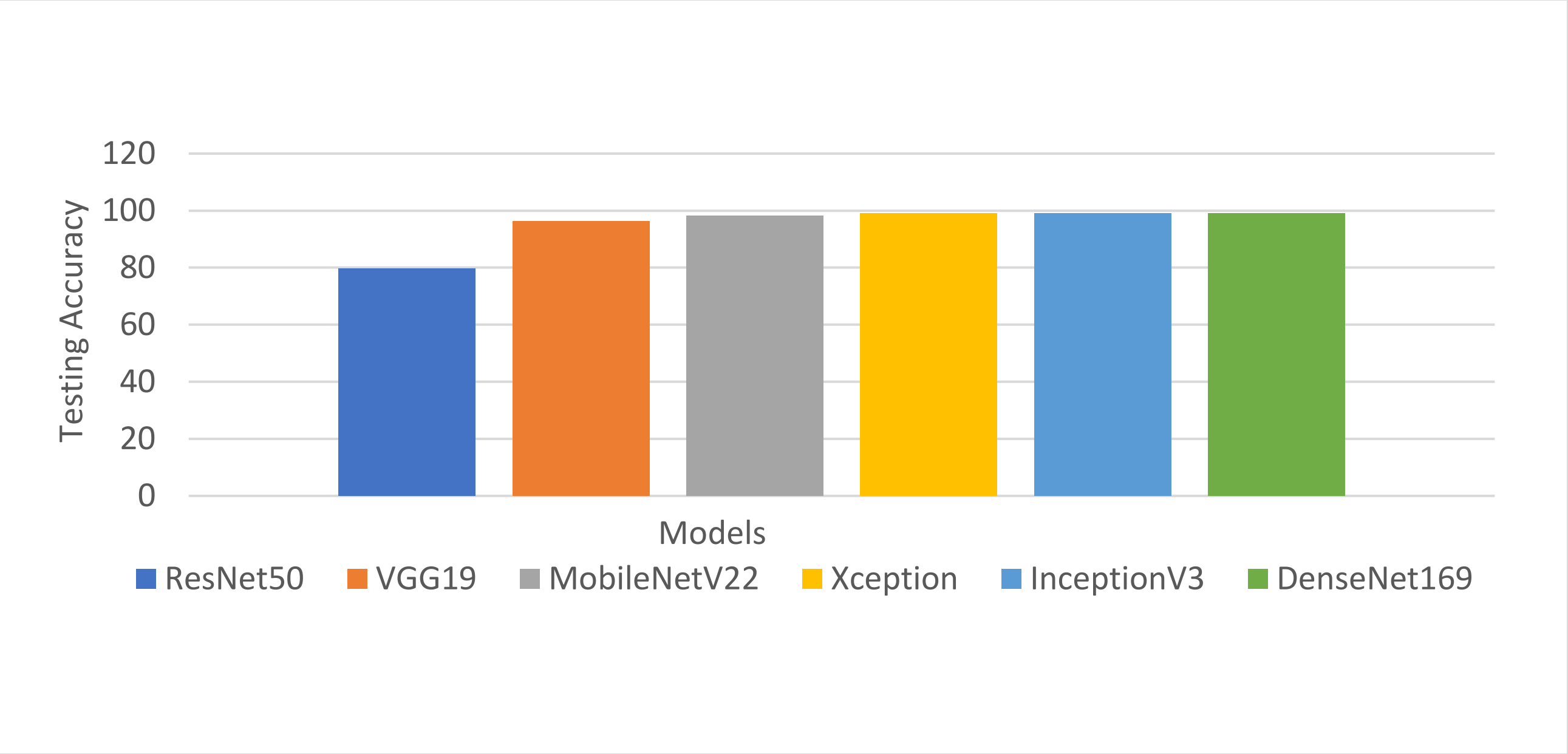

Fig.7. Comparison of testing accuracy of our models.

TABLE 7. Comparison of our work with very recent work

| Models | Sensitivity (%) | Specificity (%) | Accuracy (%) |
|---|---|---|---|
| Mohamed M. Farag et al. [23] | 98.3 | 94.55 | 97 |
| **Our proposed Model (DenseNet121)** | **100** | **100** | **100** |
| **Our Average Ensemble proposed model (DenseNet201 and DenseNet169)** | **100** | **100** | **100** |
| **Our Weighted Ensemble proposed model (DenseNet201 and DenseNet169)** | **100** | **100** | **100** |

## 5 Conclusion

Automated screening systems significantly reduce the time required to determine diagnoses, saving ophthalmologists time and money and allowing patients to be treated more quickly. Automated DR detection systems play an important role in detecting DR at an early stage. In our work, out of the individual TL architectures, DenseNet121 architecture provides the highest accuracy of 100%. Ensemble of DenseNet169 and DesneNet201 TL architectures also results 100% accuracy, 100% sensitivity and specificity. Data augmentation, parameters tuning and Global Average Pooling layer, dropout layer at bottom of pre-trained model has played a critical role in our work. An accurate determination of diabetic retinopathy in an appropriate time may help the patients to take preventive action from the very beginning. The research has some limitations. Firstly, no conventional ML classifier is used since deep learning classifier show the superiority in image classification. Secondly, Data pre-processing has been ignored. But it is important step of ML. Thirdly, only existence of DR or not has been considered here. Severity level and other symptoms of ophthalmological diseases are not taken into count. In future, this work will be carried out by considering all the limitations and will be tested in real world data in the field.

**Author contributions: MSH Talukder**: Implementation, result analysis and methodology writing; **Ak. Sarkar**: Conceptualization and drafting of manuscript; **Shramin Akter:** Literature review and introduction writing; **Nuhi-Alamin:** reviewing and correcting the manuscript.

**Funding:** This research received no external funding.

**Declaration of Competing Interest:** The authors declare that they have no conflict of interest**.**

**Data availability:** The dataset is public.